# Intrinsic Optical and Electronic Properties from Quantitative Analysis of Plasmonic Semiconductor Nanocrystal Ensemble Optical Extinction

Stephen L. Gibbs[†,§], Corey M. Staller[†,§], Ankit Agrawal[‡,∥], Robert W. Johns[∥,†], Camila A. Saez Cabezas[†], Delia J. Milliron[*,†]

[†] McKetta Department of Chemical Engineering, University of Texas at Austin, Austin, Texas, 78712-1589, United States

[‡] The Molecular Foundry, 1, Cyclotron Road, Berkeley, California 94720, United States

[∥] Department of Chemistry, University of California, Berkeley, Berkeley, California 94720, USA

[§] These authors contributed equally

*Corresponding Author
E-mail: milliron@che.utexas.edu
Telephone: (512)232-5702

ABSTRACT: The optical extinction spectra arising from localized surface plasmon resonance in doped semiconductor nanocrystals (NCs) have intensities and lineshapes determined by free charge carrier concentrations and the various mechanisms for damping the oscillation of those free carriers. However, these intrinsic properties are convoluted by heterogeneous broadening when measuring spectra of ensembles. We reveal that the traditional Drude approximation is not equipped to fit spectra from a heterogeneous ensemble of doped semiconductor NCs and produces fit results that violate Mie scattering theory. The heterogeneous ensemble Drude approximation (HEDA) model rectifies this issue by accounting for ensemble heterogeneity and near-surface depletion. The HEDA model is applied to tin-doped indium oxide NCs for a range of sizes and doping levels but we expect it can be employed for any isotropic plasmonic particles in the quasistatic regime. It captures individual NC optical properties and their contributions to the ensemble spectra thereby enabling the analysis of intrinsic NC properties from an ensemble measurement. Quality factors for the average NC in each ensemble are quantified and found to be notably higher than those of the ensemble. Carrier mobility and conductivity derived from HEDA fits matches that measured in the bulk thin film literature.

## INTRODUCTION

Optical and electronic properties of nanoparticles differ in marked ways from the properties of bulk materials. In conductive materials, a key difference is that the polarization of free charge carriers that propagates along the surface, known as a surface plasmon resonance, is spatially confined in nanoparticles. This confined surface plasmon is called a localized surface plasmon resonance (LSPR). Since the momentum conservation conditions differ from planar surface plasmons, LSPR gives rise to strong extinction of light near the resonant frequency, $\omega_{LSPR}$, which increases with the charge carrier concentration, $n_e$. Traditional plasmonic materials such as Au and Ag with intrinsic $n_e \approx 10^{22}$ have LSPR in the visible wavelengths. A less-studied class of plasmonic materials is doped semiconductor nanocrystals (NCs). These materials broaden the opportunities for plasmon-enhanced processes because their carrier concentration is relatively low and adjustable over orders of magnitude ($n_e \approx 10^{19} - 10^{21}$) simply by modifying dopant concentration.[1,2] Doped semiconductor NCs are ideal candidate materials for accessing infrared frequencies, an important wavelength regime for emerging technologies in waveguiding for telecommunication,[3–6] molecular sensing,[7–10] and photothermal theranostics.[11–13] Plasmonic enhancement of these processes relies on strong light-matter interaction within a narrow bandwidth, which can be quantified as the ratio of $\omega_{LSPR}$ to the full width at half maximum (FWHM), also known as the quality factor or Q-factor.

Factors that contribute to the FWHM of LSPR in a NC ensemble include both physical properties of the individual NCs, what we refer to as *intrinsic damping*, and the particle-to-particle variations in these properties, what we refer to as *heterogeneous broadening*. Intrinsic damping is inversely proportional to electron mobility within a NC, which is dependent on the carrier concentration and the mean free path of the charge carriers. The mean free path is proposed to be dominated by surface damping in nanostructures with one or more dimensions smaller than the bulk mean free path.[14] Size-dependent damping consistent with free electron surface damping was observed in Ag[15] and Au nanoparticles.[16] Unlike their noble metal counterparts, semiconductor NCs such as tin-doped indium oxide (ITO) form depletion regions near the surface which significantly influence LSPR.[17] These near-surface depletion layers have widths on the order of a nanometer, have significantly reduced free carrier concentration, and effectively shrink the volume accessible to conduction band electrons.[18,19] We expect the extent of depletion to be an important consideration for damping in doped semiconductors because their presence is expected to reduce the mean free path of the free charge carriers.

Typical optical models used to analyze LSPR spectra convolute intrinsic damping and heterogeneous broadening into a single damping term, leading to potential misinterpretation of material electronic properties. Despite the rather narrow size distributions achieved by recent synthetic developments in nanoparticles, size polydispersity is still often nearly 10%.[20,21] Prior work has shown that the far-field extinction spectra of doped semiconductor NCs are very sensitive to NC diameter.[22,23] When surface damping is prominent, a size distribution within an ensemble of NCs causes a distribution of intra-NC electron mobility due to variations in surface damping, contributing to heterogeneous broadening. Beyond size distribution effects, dopant incorporation also varies from NC-to-NC within an ensemble, leading to significant carrier concentration polydispersity and adding to heterogeneous broadening.[24] Indeed, when absorption spectra of single aluminum-doped zinc oxide and ITO NCs were directly measured, Johns et al. found striking variability in linewidth and in absorption peak energy within ensemble populations.[25]

The typical fit procedure for extracting material properties from LSPR extinction spectra, the simple Drude approximation (SDA), does not account for the effects of near-surface depletion layers or ensemble heterogeneity. These limitations obscure interpretation of ensemble measurements and can potentially mislead efforts to develop higher Q-factor materials. Herein, we present a model that builds on the SDA for the more incisive fitting of optical spectra of NC ensembles. The heterogeneous ensemble Drude approximation (HEDA) model uses only well-known material constants and routinely measured NC physical properties to fit for NC properties that cannot be easily measured directly, those being: carrier concentration, carrier concentration polydispersity, near-surface depletion width, and bulk mean free path. By analyzing the far-field response as a sum of contributions from individual NCs, the ensemble fit enables analysis of physical properties for individual NCs within an ensemble without the laborious effort of single

NC spectroscopy. We use ITO NCs of varying dopant concentrations and sizes as a model system where fitting results can be compared to expectations from a well-established literature to establish both the validity and potential of the new analysis procedure. We find the volume-normalized extinction coefficient, as well as the Q-factor, of an average NC is significantly higher than its corresponding ensemble, mainly due to heterogeneous broadening as a result of carrier concentration heterogeneity. Neglecting ensemble heterogeneity and near-surface depletion, the conventional SDA model failed to fit our data with physically realistic parameter values consistent with Mie scattering theory. The HEDA model rectified this inconsistency. The SDA underestimates the electron mobility of individual NCs due to the convolution of intrinsic damping and heterogeneous broadening. HEDA analysis yields an electron mobility within colloidal ITO NCs ranging from 15 to 35 $cm^2V^{-1}s^{-1}$ and extracts bulk mobility values matching those reported by Hall effect measurements in ITO thin films.[26]

**METHODS**

**NC Synthesis.** ITO NCs were synthesized by modification of methods published by the Hutchison group.[20,21] A detailed explanation can be found in Reference 23. In short, NCs are synthesized via a slow injection technique of metal-oleate into oleyl alcohol. In(III)acetate and Sn(IV)acetate are added to oleic acid and the solution is degassed and heated to 150°C for at least 2 hours to generate In-oleate and Sn-oleate. The metal-oleate solution is added dropwise by syringe pump to a flask containing degassed oleyl alcohol at 290°C. NC size and dopant concentration are controlled by the metal-oleate volume injected and the ratio of metal precursors, respectively.

**Spectroscopy measurements.** ITO NC optical properties were characterized using a Bruker Vertex 70 FTIR spectrophotometer (650-4000 $cm^{-1}$) and Agilent Cary series UV-vis-NIR spectrophotometer (3031-37000 $cm^{-1}$). A detailed explanation of optical measurements can be found in Reference 23. In short, dilute dispersions of ITO NCs in 1.8 mM oleic acid in TCE were prepared from stock solutions. Four samples of varying NC concentrations were measured in a KBr liquid cell with 0.5 mm path length for each sample in the ITO doping and size series.

**Fitting Procedure.** Optical spectra were fit using both the SDA and HEDA models. HEDA model fits were solved for using the MATLAB® code shown in SI Text 1. All four dilutions were fit independently for each sample and used to create error bars for fit variables. For each dilution, the volume fraction and NC size distribution were fixed to the measured values. The model then uses a least squares function to fit for four variables (described below) within fit constraints. Final fit values were found to be independent of initial guess values.

**RESULTS AND DISCUSSION**

**Fitting NC Ensemble Optical Spectra**

To test the robustness of this model, we compared fit quality of the HEDA and SDA models across fifteen independently synthesized samples of ITO NCs with wide-ranging doping level and size: 0 to 7.5 at% Sn and 6 to 20 nm in diameter, respectively. Details of NC synthesis and characterization are reported in Reference 23. The SDA model requires the input of pathlength and material constants and then fits for NC volume fraction in solution, $f_V$, damping constant, $\Gamma$, and plasma frequency, $\omega_p$. These fitting procedures often yield a volume fraction that doesn't match the measured value, but instead erroneously acts as a correction factor to scale the fitted extinction intensity. To eliminate this artificial scaling factor, LSPR spectra were also fit using the SDA with $f_V$ fixed to the independently measured value. When $f_V$ is fixed, the SDA is unable to simultaneously fit peak intensity and lineshape (Figure 1). We hypothesized that these discrepancies arise because the SDA uses a single damping value and a single plasma frequency value to fit an ensemble spectrum that has a distribution of those values.[27] Hereafter, only the SDA with floating volume fraction will be discussed as it is the common method for fitting optical extinction spectra. Across a wide range of doping levels and sizes, the HEDA model, unlike the SDA, reliably fits ensemble spectra with the volume fraction *fixed* to the measured value (Figure S1).

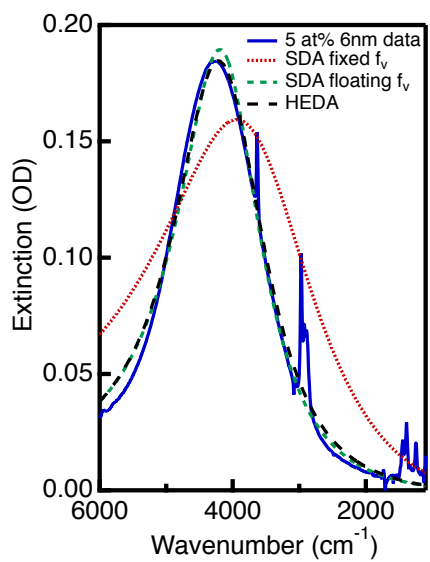

**Figure 1. Three models for fitting extinction spectra.** Simple Drude approximation (SDA) with a floating NC volume fraction $f_V$, SDA with $f_V$ fixed to the measured value, and heterogeneous ensemble Drude approximation (HEDA) fits to extinction data for 5 at% Sn 6 nm ITO NCs.

Unlike the SDA, the HEDA model accounts for ensemble heterogeneity by considering that there are finite distributions for both the NC size and carrier concentration. Beyond ensemble heterogeneity, the HEDA model improves upon the SDA by explicitly considering the impact of near-surface depletion layers and surface scattering on the optical response. To do this, in addition to the basic inputs required for the SDA model, the HEDA model takes as fixed inputs the independently measured mean NC radius, NC radius standard deviation, and $f_V$. These parameters are routinely measured, though this information is not typically used in SDA fitting procedures. The model then fits for four variables: the average carrier concentration, $n_e$, the standard deviation in carrier concentration, $\sigma_{n_e}$, the electron accessible volume fraction considering surface depletion, $f_e$, and the mean free path of an electron in the bulk material, $l_{bulk}$. The effects of NC radius and each HEDA model parameter on the extinction spectra are calculated and plotted in Figure S2. Below we outline the how the HEDA model incorporates each of these physical phenomena.

i) Near-surface depletion regions

Metal oxide surfaces are commonly passivated by adsorbed water species, including surface hydroxyls, which create a density of electronic states near the NC surface. When these surface states are below the NC bulk Fermi level they cause a decreased carrier concentration near the NC surface, known as a depletion region (Figure 2a).[17–19] Due to the buildup of electrostatic potential, depletion regions near the NC surface decrease the fraction of the NC volume accessible to mobile charge carriers as evidenced by decreased conductivity in NC films when depletion regions are prevalent.[28] The HEDA model accounts for near-surface depletion by fitting for an electron accessible volume fraction, $f_e$. The radius of the spherical volume accessible to mobile charges is then $f_e^{1/3} r_{NC}$, where $r_{NC}$ is the NC radius. For ITO, the mobile charges are electrons and this decreased radius is the electron accessible radius. The near-surface depletion creates a pseudo-core-shell geometry where the NC can be described as an electron-rich core with an electron-deficient shell. This geometry requires a modification to the dielectric function of these materials to successfully model their optical response.[18] The Maxwell-Garnett effective medium approximation (EMA) is used to define the dielectric function of a core-shell NC, $\varepsilon_{cs}$, as

$$\varepsilon_{cs}(\omega) = \varepsilon_{shell}\left(\frac{(\varepsilon_{NC}+2\varepsilon_{shell})+2f_e(\varepsilon_{NC}-\varepsilon_{shell})}{(\varepsilon_{NC}+2\varepsilon_{shell})-f_e(\varepsilon_{NC}-\varepsilon_{shell})}\right) \qquad \text{Equation 1}$$

where $\varepsilon_{shell}(\omega)$ is the dielectric function of the depleted shell and $\varepsilon_{NC}(\omega)$ is the dielectric function of the core. The EMA is applicable for variable thickness of the depleted shell, converging to $\varepsilon_{NC}(\omega)$ when $f_e = 1$.

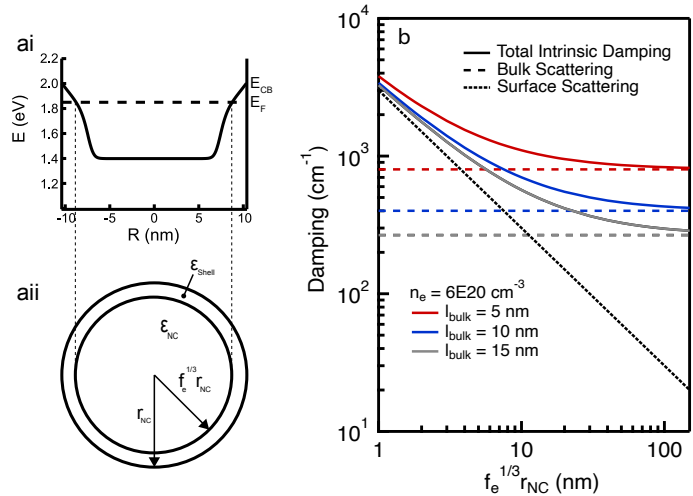

**Figure 2. Theoretical concepts for surface depletion and surface scattering.** Conduction band profile (ai) and schematic (aii) for a NC with surface depletion. Calculated total intrinsic damping and its contributions from bulk and surface scattering as a function of electron accessible radius $f_e^{1/3} r_{NC}$ for $l_{bulk}$ = 5, 10, & 15 nm at $n_e = 6 \times 10^{20}$ cm$^{-3}$ (b).

ii) Intrinsic damping

The LSPR linewidth of a single spherical NC depends on the damping constant, which is the rate at which conduction electrons scatter. The damping constant, $\Gamma$, is often used as a fitting parameter; however, free carrier damping can be calculated directly from the Drude conductivity as[16,29]

$$\Gamma = \frac{(3\pi^2)^{\frac{1}{3}}\hbar}{m_e^*} n_e^{\frac{1}{3}} \left(\frac{1}{l_{MFP}}\right) \qquad \text{Equation 2}$$

where $\hbar$ is Planck's constant, $m_e^*$ is the effective electron mass, $n_e$ is the electron concentration, and $l_{MFP}$ is the electron mean free path. For NCs of radius comparable to or less than the material's bulk mean free path, surface scattering influences the overall mean free path of NC conduction electrons (Figure 2b).[15,16] Assuming surface scattering to be specular and applying Matthiessen's rule, the electron mean free path is described by[14,30]

$$\frac{1}{l_{MFP}} = \left(\frac{1}{\frac{4}{3}r_{NC}f_e^{\frac{1}{3}}} + \frac{1}{l_{bulk}}\right) \qquad \text{Equation 3}$$

where $l_{bulk}$ is the mean free path for the bulk material. There are many reports that measure the bulk mean free path of electrons in ITO to be anywhere from 5 to 17 nm and its dependence on electron concentration is unclear.[26,31–33] Due to this uncertainty, $l_{bulk}$ is used as a fit parameter to capture its relationship to other physical parameters. Based on our search, an upper bound of 17 nm was placed on the mean free path to ensure meaningful output values and prevent small, surface scattering-dominated NCs from reporting infinite values for bulk mean free path. The intrinsic damping constant can then be defined by combining Equations 2 and 3 as

$$\Gamma = \frac{(3\pi^2)^{\frac{1}{3}}\hbar}{m_e^*} n_e^{\frac{1}{3}} \left(\frac{1}{\frac{4}{3}r_{NC}f_e^{\frac{1}{3}}} + \frac{1}{l_{bulk}}\right) \qquad \text{Equation 4}$$

Intrinsic damping is particularly size-dependent when surface damping dominates bulk damping, i.e., when the electron accessible radius is smaller than $l_{bulk}$ (Figure 2b). For larger particles, it converges to the bulk damping value above a radius about 10 times $l_{bulk}$, however this is much larger than the NC radii used in this study. With $l_{bulk}$ expected to lie between 5 and 17 nm, and the radii of the NCs in this study ranging from 3 to 10 nm, we expected surface damping to be a significant, if not dominant, factor in overall intrinsic damping.

    iii)    Ensemble heterogeneity

Size heterogeneity and electron concentration heterogeneity are modeled as Gaussian distributions. To do this, we construct a two-parameter probability density function (PDF) that ranges from the mean value plus or minus three standard deviations in NC radius and electron concentration. The PDF is then discretized with a 41x41 data point mesh containing 1681 permutations of NC radius and electron concentration. Fit results varied by less than 0.5% when the data point mesh size was changed from 31x31 to 41x41, indicating additional PDF resolution does not significantly change results (Figure S3). The total extinction of an ensemble is a probability-weighted sum of these 1681 NCs. To obtain the extinction for a two-dimensional matrix of NCs we first calculate the plasma frequency and damping of each NC, with electron accessible radius $r_{NC_i}f_e^{\frac{1}{3}}$ and electron concentration $n_{e_j}$.

$$\omega_{p_j} = \sqrt{\frac{q^2 n_{e_j}}{\varepsilon_0 m_e^*}} \qquad \text{Equation 5}$$

$$\Gamma_{ij} = \frac{(3\pi^2)^{\frac{1}{3}}\hbar}{m_e^*} n_{e_j}^{\frac{1}{3}} \left( \frac{1}{\frac{4}{3}r_{NC_i}f_e^{\frac{1}{3}}} + \frac{1}{l_{bulk}} \right) \qquad \text{Equation 6}$$

where $q$ is the electron charge and $\varepsilon_0$ is the permittivity of vacuum. The complex dielectric function $\varepsilon_{NC_{ij}}(\omega)$ is expressed using the Drude-Lorentz model

$$\varepsilon_{NC_{ij}}(\omega) = \varepsilon_\infty - \frac{\omega_{p_j}^2}{\omega^2 + i\omega\Gamma_{ij}} \qquad \text{Equation 7}$$

where $\varepsilon_\infty$ is the high-frequency dielectric constant for a given material. As described above, the presence of a depletion layer necessitates the use of a core-shell geometry for the complex dielectric function.

$$\varepsilon_{cs_{ij}}(\omega) = \varepsilon_{shell} \left( \frac{\left(\varepsilon_{NC_{ij}} + 2\varepsilon_{shell}\right) + 2f_e\left(\varepsilon_{NC_{ij}} - \varepsilon_{shell}\right)}{\left(\varepsilon_{NC_{ij}} + 2\varepsilon_{shell}\right) - f_e\left(\varepsilon_{NC_{ij}} - \varepsilon_{shell}\right)} \right) \qquad \text{Equation 8}$$

In systems of non-interacting spheres, the absorption cross-section of a single particle, $\sigma_{abs_{ij}}$, is defined by Mie theory as

$$\sigma_{abs_{ij}}(\omega) = 8\pi^2 r_{NC_i}^3 \omega \sqrt{\varepsilon_m} Imag\left\{ \frac{\varepsilon_{cs_{ij}}(\omega) - \varepsilon_m}{\varepsilon_{cs_{ij}}(\omega) + 2\varepsilon_m} \right\} \qquad \text{Equation 9}$$

where $\varepsilon_m$ is the dielectric constant of the surrounding medium. For NCs smaller than 5% of the wavelength of incident light, scattering of incident light is negligible and therefore extinction is equal to absorption.[14] This assumption holds up to at least 150 nm diameter for ITO NCs. The absorption cross-section for each of the 1681 points is probability-weighted and summed to give the effective absorption cross-section for the ensemble, $\sigma_{abs}^{eff}$, as

$$\sigma_{abs}^{eff} = \sum_i^m \sum_j^n \left( \sigma_{abs_{ij}}(\omega) p_{n_{e_j}} p_{r_{NC_i}} \Delta n_e \Delta r_{NC} \right) \qquad \text{Equation 10}$$

where $p_{n_{e_j}}$ and $p_{r_{NC_i}}$ are the probabilities of $n_{e_j}$ and $r_{NC_i}$, respectively, $\Delta n_e$ and $\Delta r_{NC_i}$ are the step sizes for $n_e$ and $r_{NC}$, respectively, and $m$ and $n$ are the mesh dimensions (41 here). The effective absorption cross-section of the ensemble is then plugged into the Beer-Lambert law,

$$A = \frac{f_v l}{\ln(10) V} \sigma_{abs}^{eff} \qquad \text{Equation 11}$$

where $V$ is the average volume of a NC, defined as

$$V = \sum_i^m \sum_j^n \left( \frac{4}{3} \pi r_{NC_i}^3 p_{n_{e_j}} p_{r_{NC_i}} \Delta n_e \Delta r_{NC} \right) \qquad \text{Equation 12}$$

The HEDA model fitting procedure was demonstrated to be robust by using a variety of initial guesses for each fitted parameter. Changing the initial guesses within fit parameter constraints did not change the solution (Table S1). Aside from fitting errors, one potential source of error for the HEDA model is non-physical fit parameter correlations. Our model contains multiple parameters that influence the FWHM (specifically $\sigma_{n_e}$, $f_e$, and $l_{bulk}$) and so we plotted those parameters against each other and found no correlations (Figure S4). Independence of all broadening factors among a diverse set of samples supports that the HEDA model yields meaningful values and is a robust model. We note that fitting for a single $f_e$ and $l_{bulk}$ rather than distributions of these parameters was an appropriate simplification because the heterogeneity in these parameters is expected to be much smaller than in NC radius or carrier concentration.

**HEDA Fit Results**

The 15 investigated samples can be split into two series: a doping series from 0 to 7.5 at% Sn at 20 nm diameter and a size series from 6 to 20 nm diameter at 5 at% Sn. To account for any deviation occurring as a result of colloid preparation or spectra collection, four spectra were collected for each sample from independently prepared dilute dispersions of NCs in tetrachloroethylene (TCE). The plots and error bars illustrate the average and standard deviation of the parameters extracted from the four measurements. Sample details and fit results are summarized for the NC size (Table S2) and dopant concentration (Table S3) series. Statistically significant correlations were determined by t-test against the null hypothesis of zero correlation between variables, i.e., the slope is zero. The null hypothesis was rejected for those variables with significance values of $\alpha < 0.05$, confidently concluding the existence of a relationship between them (Tables S4 and S5).

    i)       Quantifying ensemble heterogeneity

The polydispersity $D$ (defined here as the standard deviation divided by the average value in a Gaussian distribution) in radius was quantified from small angle X-ray scattering (SAXS). There was a decreasing trend in $D_R$ with radius (Figure 3a) and increasing trend in $D_R$ with dopant concentration (Figure 3b). The trends in size elucidate that under our synthetic conditions we can achieve the lowest polydispersity in larger, lower-doped NCs. While $D_R$ was measured by SAXS, the polydispersity in carrier concentration, $D_{n_e}$, was determined from the HEDA fitting. In the ideal case, dopant incorporation varies minimally within an ensemble and follows Poissonian statistics.[24,27] At high dopant incorporation, the standard deviation for number of dopant atoms incorporated in single NCs, $N_d$, is approximated by Poissonian statistics as $\sigma_{N_d} \approx \sqrt{\mu_{N_d}}$, where $\mu_{N_d}$ is the average value of $N_d$ in the NC ensemble.[25] Converting this expectation into carrier concentration, Poissonian statistics predict a drop in carrier concentration polydispersity with increasing radius to a power law dependence $D_{n_e} \propto r^{-1.5}$ and with increasing donor density $D_{n_e} \propto N_d^{-0.5}$, plotted as the solid red lines in Figures 3a and 3b. Our results contradicted this expectation, showing no trend in $D_{n_e}$ with either $r_{NC}$ or $N_d$. Moreover, $D_{n_e}$ was significantly larger across all sizes and dopant concentrations than the Poissonian expectation, consistent with the elevated dopant heterogeneity previously reported based on single particle measurements.[25] Reducing dopant heterogeneity to the Poissonian limit presents an opportunity for synthetic development to advance the utility of doped semiconductor NCs and improve ensemble optical performance. For example, peak extinction of a NC ensemble with average values $r_{NC}$ = 10 nm and $n_e$ = 6x10$^{20}$ cm$^{-3}$ would be increased by 17% when the $D_{n_e}$ drops from 10% to the Poisson limit (2%).

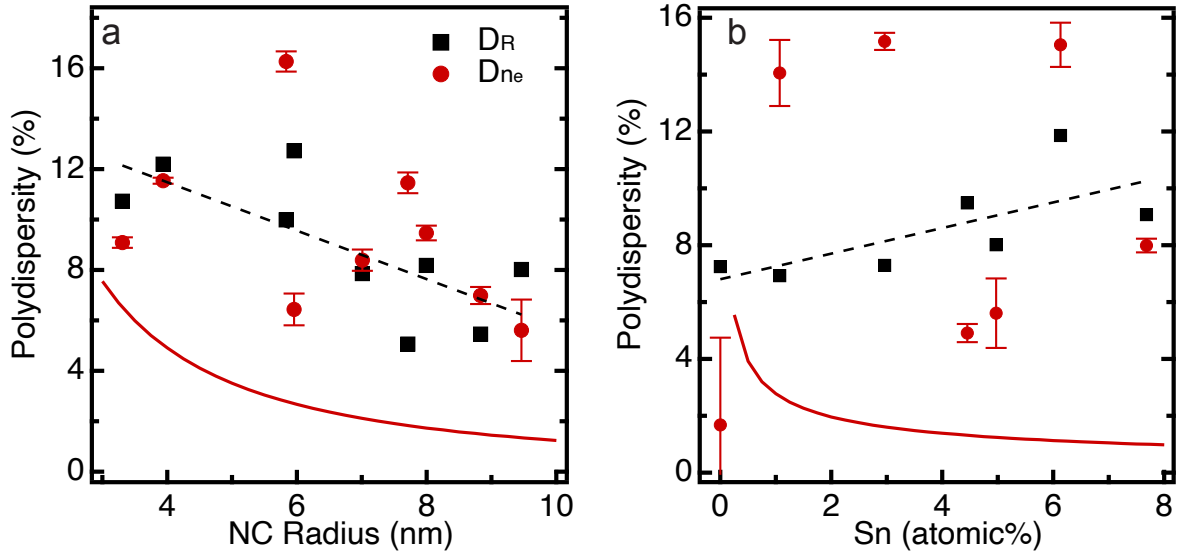

**Figure 3. Quantification of ensemble heterogeneity.** Polydispersity in radius $D_R$ (black squares) measured by small angle X-ray scattering (SAXS) and polydispersity in carrier concentration $D_{n_e}$ (red circles) extracted by fitting with the HEDA model as a function of NC radius (a) and at% Sn (b). Dashed lines show linear fits to each sample set with statistically significant correlation. Red, solid lines illustrate the expected polydispersity for dopant incorporation dictated by Poissonian statistics. Error bars indicate standard deviation in the results from HEDA fitting for four independently prepared dispersions of each sample.

    ii)       Intrinsic properties for the average NC in an ensemble

Fitting ensemble spectra with the HEDA model enables extraction of intrinsic material properties distinguished from the convoluting effects of heterogeneous peak broadening. Application of the HEDA model to a range of ITO NCs precisely tracks the trends in fit parameters with NC size and doping. Electron concentration, $n_e$, increases with dopant concentration (Figure 4aii), but does so more slowly above 4.5 at% Sn due to decreasing dopant activation (Figure S5), which is defined as the ratio between $n_e$ and the measured dopant concentration. Dopant activation is expected to decrease at higher dopant concentrations and increase with larger NC radius.[23,34,35] While dopant activation did increase with NC radius (Figure S5), there was no correlation between $n_e$ and NC radius (Figure 4ai). We suspect this is due to the moderate dopant concentration variations between samples in the size series (Table S2) that could obscure a trend in $n_e$. Electron accessible volume fraction, $f_e$, increases with NC radius and dopant concentration (Figures 4ai and 4aii), in agreement with prior work.[17,18,36,37] The near-surface depletion region narrows with increased NC size due to decreased surface area to volume ratio and with a higher concentration of ionized donors that screen the surface potential offset.

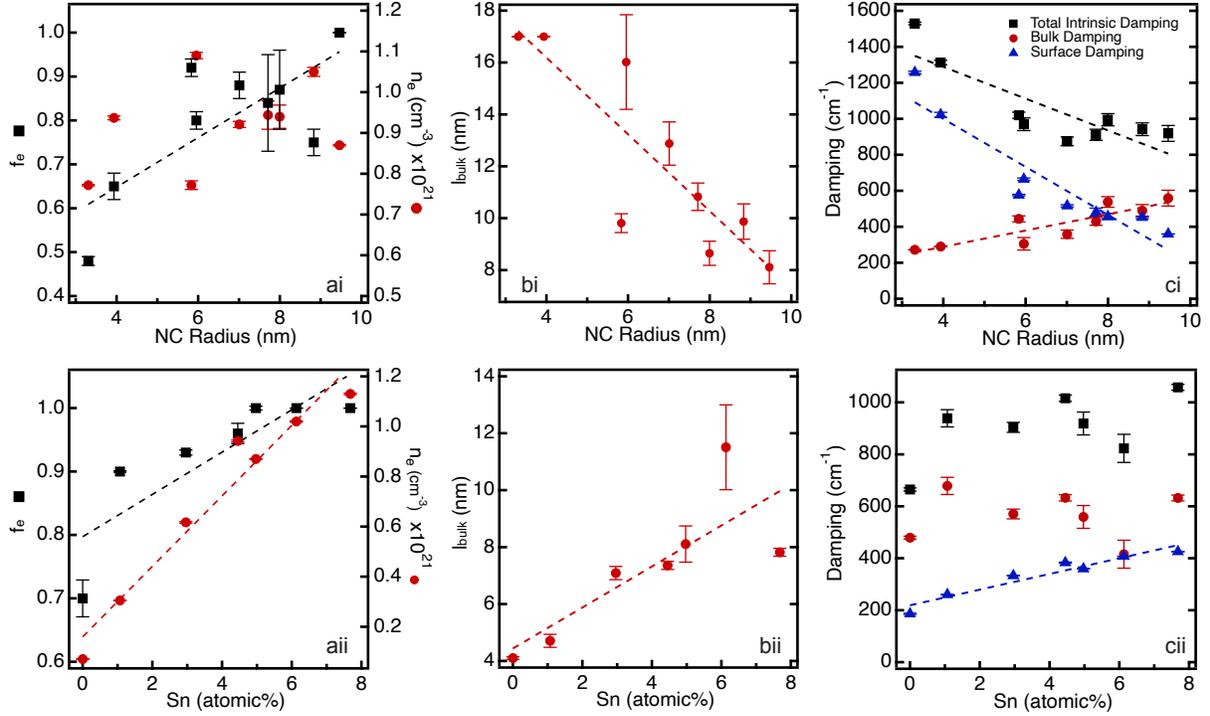

**Figure 4. HEDA Fit results.** Electron accessible volume fraction and free charge carrier concentration as a function of NC radius (ai) and doping level (aii). Bulk mean free path as a function of NC radius (bi) and doping level (bii). Total intrinsic damping as well as its substituent components, bulk and surface damping, as a function of NC radius (ci) and doping level (cii). Dashed lines show linear fits to each sample set with statistically significant correlation. Error bars indicate standard deviation in the results from HEDA fitting for four independently prepared dispersions of each sample.

The bulk mean free path, $l_{bulk}$, decreases with increasing radius (Figure 4bi). We expect the bulk damping in ITO of this dopant concentration to be dominated by electron-phonon scattering,[26] so this trend may suggest a size-dependence in the phonon behavior. However, the dominance of surface over bulk damping for smaller NCs makes this determination tentative based on the present analysis. Increasing $l_{bulk}$ with $n_e$ (Figure 4bii) has been theoretically predicted,[33] but due to the difficulty in deconvoluting the various damping mechanisms, most experimental work only reports variations in aggregate carrier damping or mobility with Sn content but does not calculate mean free path.[26] Theoretically, increasing $n_e$ will increase the Fermi velocity of the most energetic electrons but will also increase the frequency of electron scattering events, leaving an unclear prediction for mean free path. Our experimental results suggest the dominant effect of increasing $n_e$ is to increase the mean free path.

These fit results combine to determine the trends in intrinsic damping for ITO NCs. Intrinsic damping reflects the frequency of free electron scattering events in a single NC and can be categorized into two contributions: bulk damping, $\Gamma_{bulk}$, and surface damping, $\Gamma_{surface}$. From Equation 4, we define bulk and surface damping as

$$\Gamma_{bulk} = \frac{(3\pi^2)^{\frac{1}{3}}\hbar}{m_e^*} n_e^{\frac{1}{3}} \left(\frac{1}{l_{bulk}}\right) \qquad \text{Equation 13}$$

$$\Gamma_{surface} = \frac{(3\pi^2)^{\frac{1}{3}}\hbar}{m_e^*} n_e^{\frac{1}{3}} \left(\frac{1}{\frac{4}{3}r_{NC}f_e^{\frac{1}{3}}}\right) \qquad \text{Equation 14}$$

Intrinsic damping sharply drops with increasing size up to a radius of ~6 nm, above which damping becomes less size-dependent (Figure 4ci). This trend results from the reduction in $\Gamma_{surface}$, even with a steady increase in $\Gamma_{bulk}$. In fact, our results show that surface damping is dominant over bulk damping for NCs smaller than ~8 nm in radius. Rising $r_{NC}$ and $f_e$ drive the reduction in $\Gamma_{surface}$, while diminishing $l_{bulk}$ contributes to the moderate increase in $\Gamma_{bulk}$. Intrinsic damping was not correlated with dopant concentration (Figure 4cii), contrary to expectations suggested by Equation 2. This invariance results from $l_{bulk}$ increasing with dopant concentration, counteracting the effects of increased $n_e$. The moderate trend in $\Gamma_{surface}$ with doping was not strong enough to drive a trend in total intrinsic damping. While $\Gamma_{surface}$ rises due to growing $n_e$, it is mitigated by increasing $f_e$. Reduction in surface damping due to a reduction in depletion layer thickness (increasing $f_e$) is a significant factor influencing the optoelectronic properties of ITO NCs and is expected to be for other degenerately doped semiconductor NCs.

By removing the contribution of heterogeneous broadening and deconvoluting surface and bulk damping, the HEDA model uncovers trends in NC material properties. It is important that the HEDA model achieves high quality fits for a range of ITO NCs without incorporating frequency-dependent damping, as previously included to modify the simple Drude dielectric function.[19,34,38,39] As ionized impurity scattering is expected to be frequency-dependent, the present success in fitting supports the idea that impurity scattering is not a dominant scattering source for ITO within this doping and size range.

**HEDA Fit Results Align with Mie Scattering Theory**

While the extinction coefficient is commonly reported on a molar (i.e., number-normalized) basis, it is helpful to report the volume-normalized basis to examine size-dependence of intrinsic properties. The volume-normalized extinction coefficient, $\epsilon_{NC}$, is defined as

$$\epsilon_{NC} = \frac{\sigma_{abs}}{V} \qquad \text{Equation 13}$$

A rigorous analysis of Mie theory, as shown in Reference 23, reveals that the volume-normalized extinction coefficient for a single spherical particle in the quasistatic regime and with $\omega_p \gg \Gamma$ – both conditions that are valid for the ITO NCs in this work – can be well-approximated as

$$\epsilon_{NC} = \frac{18\pi \varepsilon_m^{\frac{3}{2}}}{(\varepsilon_\infty + 2\varepsilon_m)} \left(\frac{\omega_p^2}{\Gamma(\varepsilon_\infty + 2\varepsilon_m)}\right) \qquad \text{Equation 14}$$

From this we can calculate the expected extinction of a single NC *a priori* given the values of $\omega_p$ and $\Gamma$. We note that, in the absence of near-surface depletion regions, extinction increases linearly with $\frac{\omega_p^2}{\Gamma}$; however, when considering the effect of depletion, extinction increases linearly with $f_e \frac{\omega_p^2}{\Gamma}$. This extinction coefficient dependence can be used to check the validity of an optical absorption fitting procedure. The volume-normalized extinction for the ensemble measurements and fit results for all samples are plotted in Figures S6 and S7.

The volume-normalized extinction coefficients extracted from the traditional SDA model with floating volume fraction compare poorly to the theoretically predicted relationship (Figure 5a). The disagreement indicates that the extracted fit values are not reconcilable with the expected peak extinction coefficient according to Mie theory. Near-surface depletion and ensemble heterogeneity reduce the volume-normalized extinction observed in ensemble measurements such that the SDA is insufficient to describe the essential physics underlying the LSPR. The SDA is able to simultaneously fit experimental data and yet violate Mie theory by floating $f_v$ to non-physical values, arbitrarily scaling intensity.

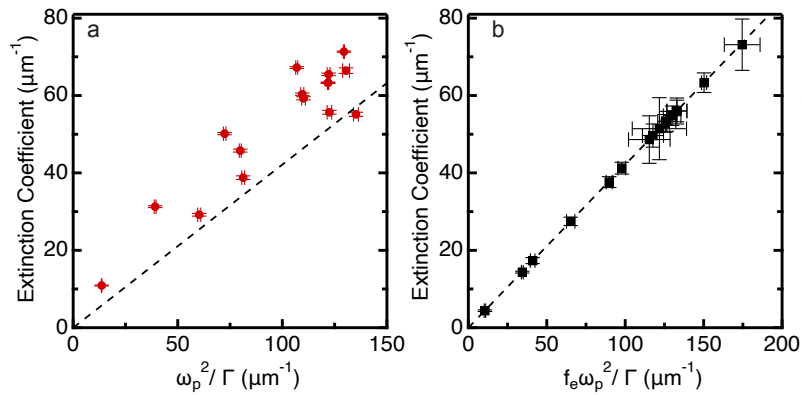

**Figure 5. Ensemble and average NC extinction.** Volume-normalized peak extinction coefficient extracted from SDA (a) and HEDA (b) fits to ensemble spectra. The dashed line indicates the expected extinction coefficient derived from Mie scattering theory. Error bars indicate standard deviation in the results from HEDA fitting for four independently prepared dispersions of each sample.

When calculating peak extinction from simulated spectra (Figure S2), the theoretical linear trend is found for all values of $r_{NC}$, $n_e$, and $l_{bulk}$, but incorporation of $D_{n_e}$ or $f_e$ violated the trend (Figure S8). The deviation from Mie theory for varying depletion layer thickness was rectified by normalizing by $f_e$ so that only the volume accessible to free electrons is considered. These calculations emphasize that, in particular, the ability of the HEDA model to account for $D_{n_e}$ and $f_e$ allows it to fit an ensemble spectrum without sacrificing the validity of intrinsic material properties or violating the expectations from Mie scattering theory. For these reasons, when using HEDA fit parameters to calculate the volume-normalized extinction coefficient for an average (representative) NC in each ensemble, the peak extinction values fall directly in line with theory (Figure 5b). The success of the HEDA model in uncovering intrinsic NC properties as well as the effect of ensemble heterogeneity can elucidate the critical parameters for improving key optical and electronic properties such as Q-factor and carrier mobility.

**Optical and Electronic Properties Derived from HEDA Model Fit Results**

The Q-factor of a NC ensemble not only depends on intrinsic NC properties, but also on ensemble heterogeneity. The HEDA model diagnoses how to improve the optical performance for a batch of NCs. Because damping significantly decreases with increasing NC radius and $\omega_{LSPR}$ blue-shifts with higher doping, it follows that the Q-factor (a ratio of $\omega_{LSPR}$ to FWHM) increases with both radius and doping (Figures 6a and 6b). In agreement with prior work, Q-factors for ensemble spectra ($Q_{ensemble}$) spanned the range of 2.0 to 4.6.[40] Using parameters derived from the HEDA model, the Q-factor for the average NC within each ensemble ($Q_{NC}$) was calculated and compared to $Q_{ensemble}$. $Q_{NC}$ was notably higher than $Q_{ensemble}$ for nearly all samples, with the maximum ratio reaching 1.8-fold for a 6 at% Sn 20 nm ITO NC with $Q_{NC}$ of 6.2. To our knowledge 6.2 is higher than any reports of $Q_{ensemble}$ for ITO NCs. The ratio of $Q_{NC}$ to $Q_{ensemble}$ strongly correlates with $D_{n_e}$ (Figure 6c), emphasizing the potential enhancements to ensemble quality factor achievable when narrowing the distribution in $n_e$. While a perfectly uniform ensemble of doped NCs is unachievable, the rise in Q-factor across nearly all samples motivates the pursuit of reducing ensemble heterogeneity.

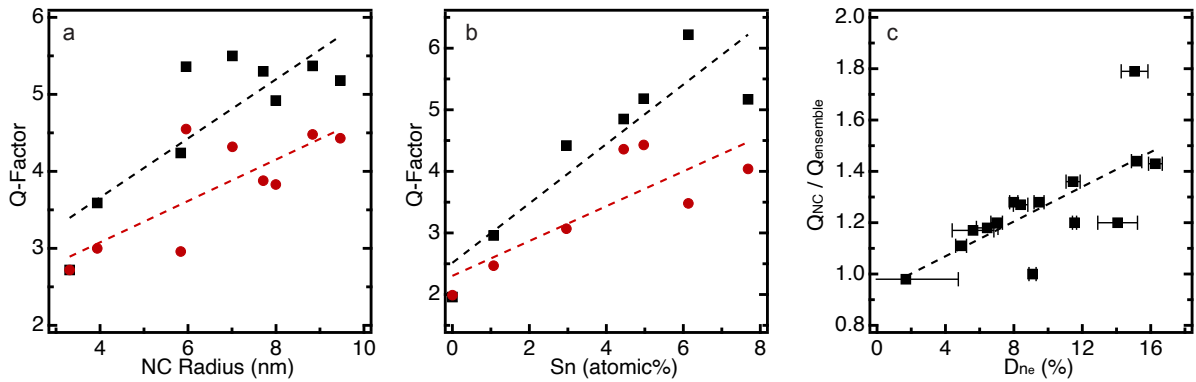

**Figure 6. Ensemble and average NC optical properties.** Q-factor calculated from the ensemble spectrum ($Q_{ensemble}$, red circles) and the spectrum of the average NC within that ensemble acquired through HEDA modeling ($Q_{NC}$, black squares) as a function of NC radius (a) and doping level (b). Ratio of $Q_{NC}$ to $Q_{ensemble}$ as a function of carrier concentration polydispersity ($D_{n_e}$) for all ITO NC samples. Dashed lines show linear fits to each sample set with statistically significant correlation. Error bars indicate standard deviation in the results from HEDA fitting for four independently prepared dispersions of each sample.

In parallel to optical properties, quantification of electronic properties of interest, such as mobility and conductivity, benefit from decoupling the intrinsic properties from heterogeneous broadening. While the electronic properties of thin films are readily measured and understood, these properties are significantly more complicated to rationalize and define for films comprised of NCs; optical analysis of intrinsic properties can be significantly enabling for such analyses. For example, although inter-NC charge transfer is a key bottleneck in achieving high electronic performance in thin films of NCs, it is not easily measured directly.[41] Instead, inter-NC resistance in metal oxide NC films has been calculated by measuring overall film resistance and subtracting the contribution of intra-NC resistance determined using fits to optical extinction data.[22,28,42] The accuracy of calculated values for inter-NC resistance is determined by the quality of the optical fits.

The intra-NC conductivity and mobility are calculated from the electron concentration and damping constant determined by optical fitting. Once the heterogeneous contributions are eliminated, the average NC spectrum indicates the intrinsic damping is less than would be reported using a conventional SDA fitting approach. We compared damping extracted from SDA, referred to as ensemble damping ($\Gamma_{ensemble}$), to that of an average NC determined using HEDA fitting, i.e. intrinsic damping ($\Gamma$). The difference between $\Gamma_{ensemble}$ and $\Gamma$ illustrates how the extracted fit parameters can differ when heterogenous broadening and intrinsic damping are convoluted and deconvoluted, respectively (Figure 7a). The SDA gives an ensemble damping value that is, on average, 19% higher than the intrinsic damping provided by the HEDA model. The maximum disparity between damping values was 65% for the 20 nm 6.5 at% Sn ITO sample, which also had one of the highest values of carrier concentration polydispersity (Figure 3b). Further, the two cases where SDA damping and HEDA intrinsic damping are in near agreement are 20 nm 0 at% Sn ITO and 6 nm 5 at% Sn ITO, which are dominated by intrinsic damping due to low carrier concentration polydispersity and high surface damping, respectively.

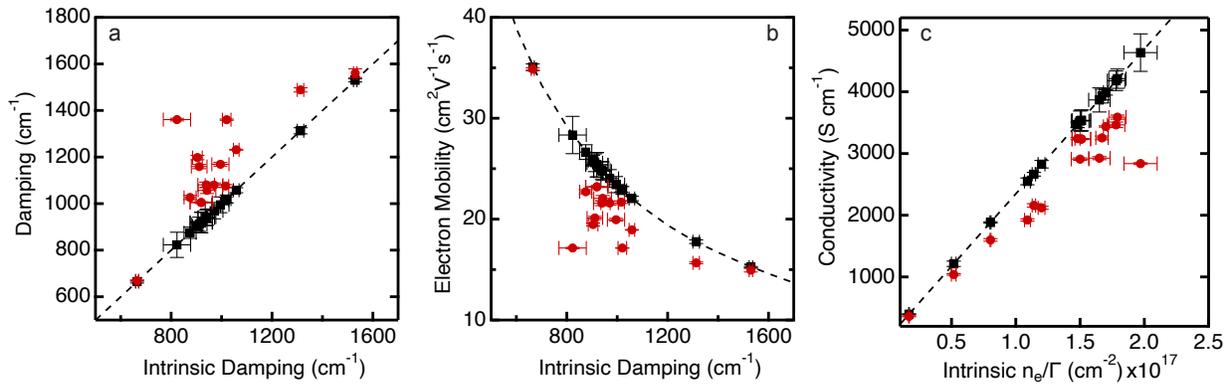

**Figure 7. Ensemble and average NC electronic properties.** Ensemble damping ($\Gamma_{ensemble}$) as fit by the SDA model with floating volume fraction (red circles) and intrinsic damping ($\Gamma$) calculated from HEDA model fit parameters (black squares) (a). Fitted damping and electron concentrations were used to calculate electron mobility (b) and conductivity (c). Dashed lines calculate the electronic parameters extrapolated beyond values extracted from our sample sets. Error bars indicate standard deviation in the results from HEDA fitting for four independently prepared dispersions of each sample.

Electron mobility is an important electronic parameter for semiconductors, describing how quickly an electron moves through a material experiencing an applied electric field. Following directly from the observed differences in damping, the SDA-derived ensemble mobility is significantly lower than the intrinsic mobility (Figure 7b). The HEDA-derived intrinsic mobility includes a significant contribution from surface damping, which is not present in ITO thin films where reported electron mobilities range from ~20 to ~80 cm$^2$/Vs. However, as described previously, the HEDA model also deconvolutes intrinsic damping into surface damping and bulk damping (Figures 4ci and 4cii). From the HEDA-derived bulk damping values, the bulk mobility calculated for these ITO NCs ranges from ~35 cm$^2$/Vs to ~85 cm$^2$/Vs, nearly identical to the range of literature values for ITO thin films with thickness much greater than $l_{bulk}$ (Figure S9a).[26,43] Correctly reproducing bulk ITO mobility provides is powerful evidence for the efficacy and accuracy of the HEDA model and its ability to deconvolute the contributions of various damping mechanisms.

Optically derived intra-NC conductivity is calculated from the Drude conductivity equation using electron mobility and electron concentration in both the ensemble and intrinsic cases (Figure 7c). The maximum ensemble intra-NC conductivity found is about 3600 S/cm. In contrast, the HEDA model derived intrinsic intra-NC conductivity has a maximum of nearly 4700 S/cm, 30% higher. Such deviations have major impacts on understanding electron conduction through NC films. For example, accurate intrinsic properties are necessary to pin down design parameters for NC films such as the critical value of inter-NC resistance, below which NC films behave as metallically conductive thin films. The bulk conductivity calculated for these NCs, setting aside the effects of surface damping, ranges from ~550 S/cm to ~13400 S/cm (Figure S9b), once again matching thin film literature.[26,43]

**CONCLUSIONS**

The standard SDA procedure for fitting optical extinction yields results that overestimate intrinsic damping and violate Mie scattering theory for even the most synthetically refined NC ensembles, which nonetheless have significant dopant heterogeneity and surface depletion. A novel fitting procedure was generated to rectify the shortcomings of the SDA model. The HEDA model was proven valid based on optical fits for a wide range of NC sizes and doping levels that produced results that (1) agree Mie Scattering theory and (2) derive electronic mobility and conductivity that match well with empirical thin film literature. The HEDA model quantifies how LSPR peak shape, position, and intensity result from contributions of depletion layer thickness, surface damping, bulk damping, as well as $n_e$ and $\Gamma$ heterogeneity – a richer trove of information when compared to obtaining only $\omega_p$ and $\Gamma$ from the SDA. With these values in hand, we rationalized trends in ITO NCs with varying dopant concentration and radius.

In the size regime investigated here, intrinsic damping has significant contributions from surface damping. Thus, the average size of the NC had a strong impact on overall damping, but with $D_{r_{NC}}$ at or below 10%, size polydispersity was not a strong contributor to heterogeneous broadening. Heterogeneous broadening mainly resulted from NC-to-NC variations in electron concentration. Our results suggest that to improve the ensemble extinction it will be more effective to narrow dopant heterogeneity. Because the polydispersity is significantly above that dictated by the Poisson limit, this is a physically realizable goal. When we quantified mean free path and subtracted the contribution of surface damping, our analysis recovers bulk electron mobility and conductivity values comparable to those measured in thin films. This agreement between electronic properties of NCs and conventional thin films indicates that colloidal synthesis produces materials of high electronic quality. We expect this model is valid for isotropic plasmonic particles within the quasistatic regime of any material class. By extrapolating this one-dimensional model into two or three dimensions, we expect anisotropic particles could also be reliably modeled and their properties analyzed.

**Supporting Information**
SDA and HEDA fits plots and results for all samples, calculated extinction spectra sweeping through all HEDA parameters, model robustness, validation of mesh size, damping parameter correlation testing, correlation assessment, dopant activation, single NC optical extinction simulations, ensemble optical extinction simulations, trends in calculated optical extinction, bulk ITO electronic properties, and MATLAB® code.

**Author Information**
Corresponding Author
E-mail: milliron@che.utexas.edu

ORCID ID

Stephen L. Gibbs: 0000-0003-2533-0957

Ankit Agrawal: 0000-0001-7311-7873

Delia J. Milliron: 0000-0002-8737-451X

Notes:
The authors declare no competing financial interest.

**Author Contributions**

C.M.S. and S.L.G. wrote and developed the HEDA model and synthesized, characterized, measured optical extinction, and fit optical extinction of ITO NCs. A.A. assisted in conceptualizing and writing the HEDA model. R.W.J. assisted in experimental design and contributed intellectually. C.A.S.C. conducted and analyzed SAXS measurements. D.J.M. provided overall guidance. C.M.S., S.L.G., and D.J.M. wrote the manuscript with critical input from all the authors.

**Acknowledgment**

This research was supported by the National Science Foundation (NSF), including NASCENT, an NSF ERC (EEC-1160494, C. Staller), CHE-1905263 (A. Agrawal), the University of Texas at Austin MRSEC (DMR-1720595, C. Saez Cabezas), a Graduate Research Fellowship under Award Number (DGE-1610403, S. Gibbs), and the Welch Foundation (F-1848, R. Johns). This work was performed at UT Austin and utilized the SAXS instrument acquired under an NSF MRI grant (CBET- 1624659) and in part at the Molecular Foundry, Lawrence Berkeley National Laboratory, which is supported by the Office of Science, Office of Basic Energy Sciences, of the U.S. Department of Energy (DOE) under Contract No. DE-AC02-05CH11231.

# Intrinsic Optical and Electronic Properties from Quantitative Analysis of Plasmonic Semiconductor Nanocrystal Ensemble Optical Extinction

Stephen L. Gibbs[†,§], Corey M. Staller[†,§], Ankit Agrawal[‡], Robert W. Johns[∥,†], Camila A. Saez Cabezas[†], Delia J. Milliron[*,†]

[†]McKetta Department of Chemical Engineering, University of Texas at Austin, Austin, Texas, 78712-1589, United States

[‡] The Molecular Foundry, 1, Cyclotron Road, Berkeley, California 94720, United States

[∥] Department of Chemistry, University of California, Berkeley, Berkeley, California 94720, USA

[§] These authors contributed equally

*Corresponding Author
E-mail: milliron@che.utexas.edu
Telephone: (512)232-5702

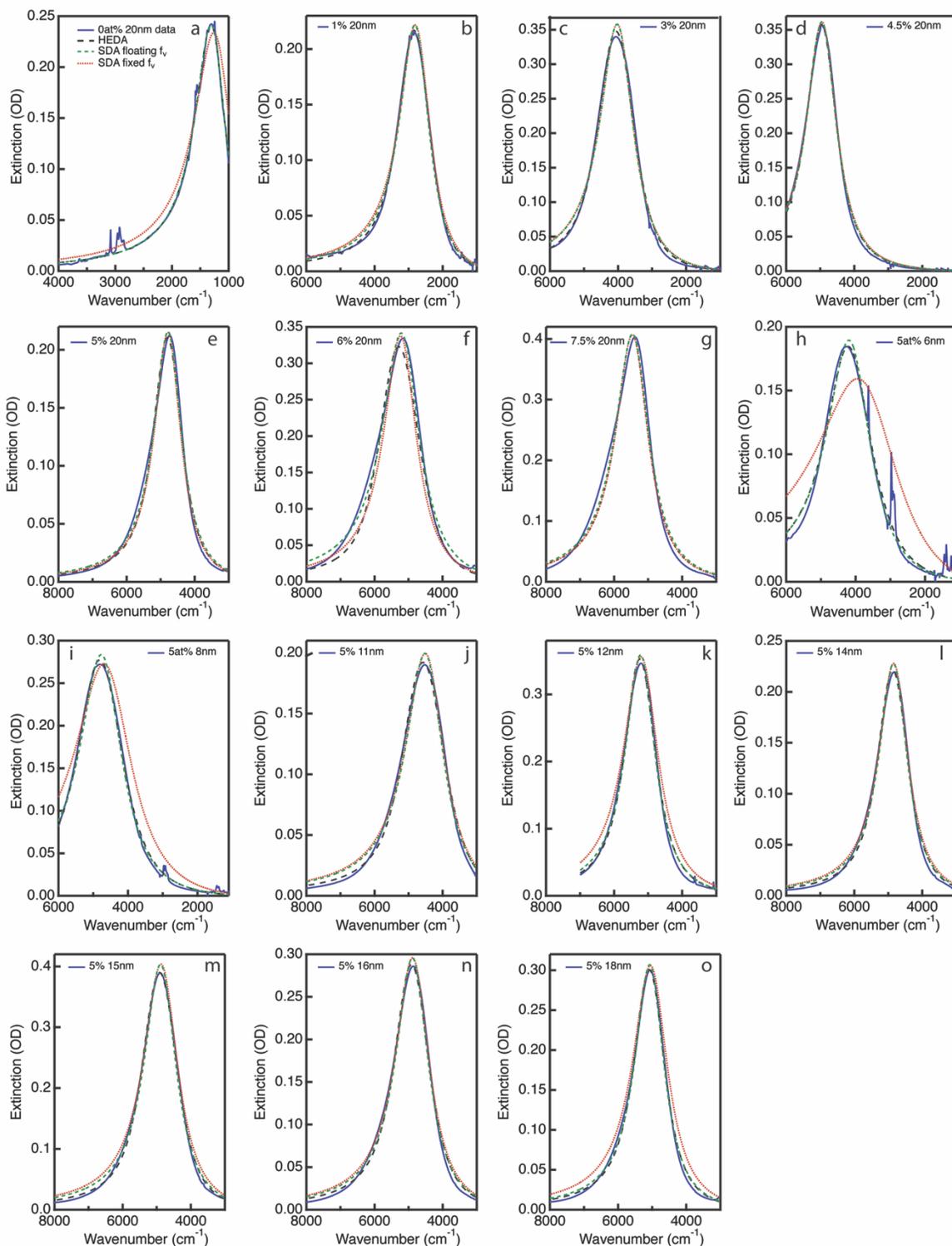

**Figure S1. Optical extinction fits** for SDA with a floating NC volume fraction, SDA with measured NC volume fraction, and HEDA model for 20 nm 0 at% (a), 1 at% (b), 3 at% (c), 4.5 at% (d), 5 at% (e), 6 at% (f), and 7.5 at% (g) ITO NCs and 6 nm (h), 8 nm (i), 11 nm (j), 12 nm (k), 14 nm (l), 15 nm (m), 16 nm (n), and 18 nm (o) 5 at% ITO NCs.

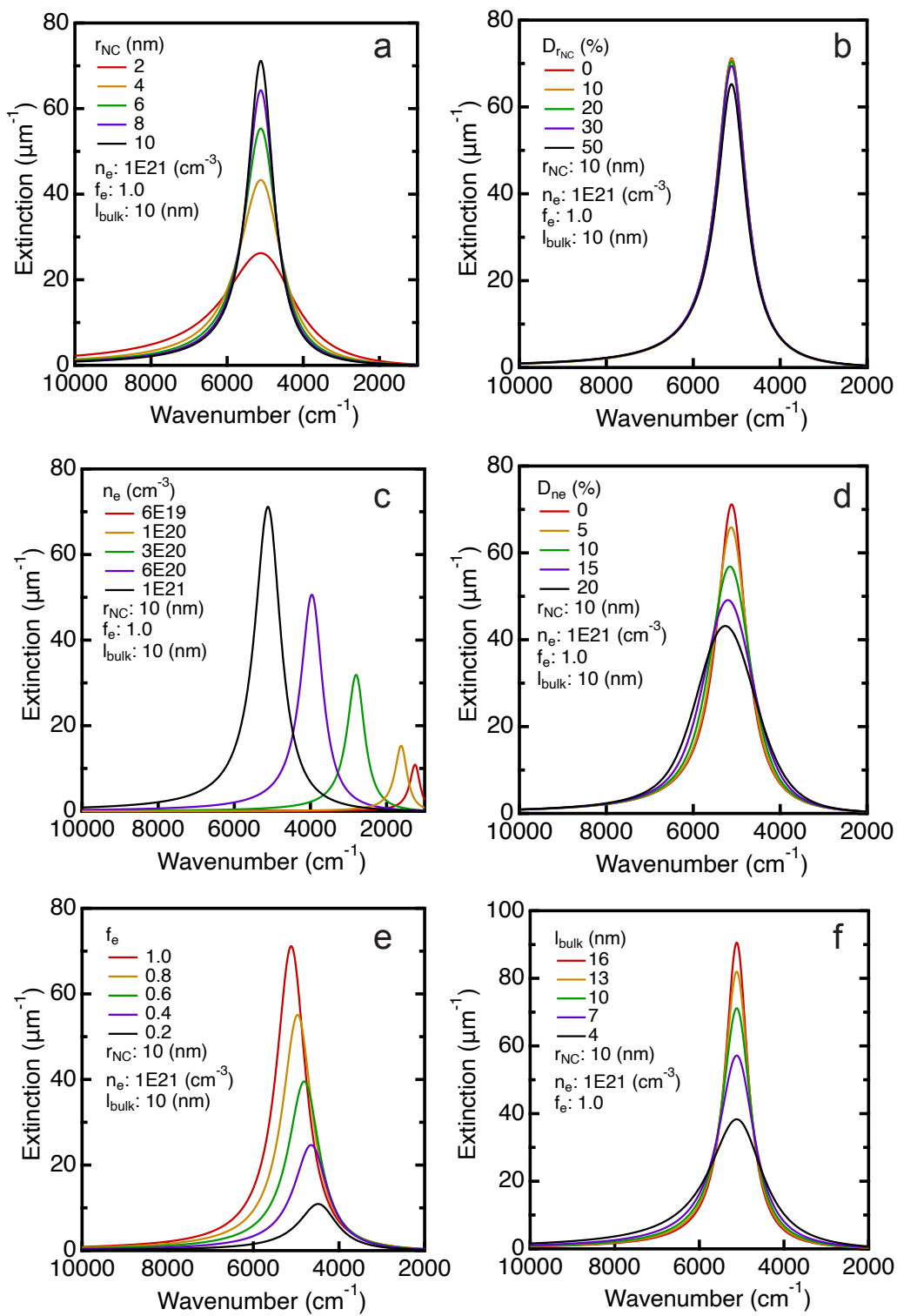

**Figure S2. Calculated LSPR spectra.** Volume-normalized extinction spectra for ITO NC ensembles of varying NC radius (a), NC radius polydispersity (b), electron concentration (c), electron concentration polydispersity, electron accessible volume fraction (e), and bulk mean free path (f).

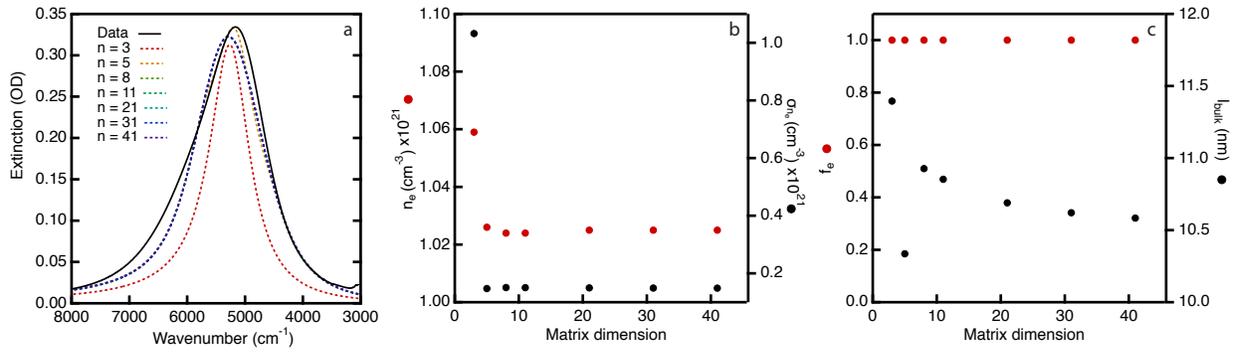

**Figure S3. Verification of mesh size.** HEDA model fits to 20 nm 6 at% ITO NCs of varying probability matrix dimensions, n, (a) and resulting variables (b) and (c). Fitted variables become more stable as n increases. Fit results varied by less than 0.5% when the data point mesh size was changed from 31x31 to 41x41. Results presented in the paper represent n=41 to ensure stabilization of fit parameters.

**Table S1. Initial guess robustness.** Eight sets of initial guesses were randomly generated between the upper and lower bounds listed below. All initial guesses were applied to fits for the 5% 11 nm diameter ITO sample spectrum and they returned exact same fit results each time.

| Fit Parameter | $n_e$ | $\sigma_{n_e}$ | $f_e$ | $l_{bulk}$ |
|---|---|---|---|---|
| Lower Bound | 1.0E+19 | 1.0E+19 | 0 | 0 |
| Upper Bound | 2.0E+21 | 2.0E+21 | 1 | 17 |
| Randomly Generated Initial Guess | 8.5E+20 | 1.8E+21 | 0.79 | 16.3 |
| | 1.4E+21 | 1.5E+21 | 0.74 | 6.7 |
| | 1.3E+21 | 3.5E+20 | 0.71 | 0.5 |
| | 5.6E+20 | 1.0E+20 | 0.10 | 14.0 |
| | 1.4E+21 | 6.4E+20 | 0.95 | 0.6 |
| | 3.8E+20 | 9.8E+20 | 0.45 | 11.0 |
| | 1.4E+21 | 1.5E+21 | 0.28 | 11.6 |
| | 1.3E+21 | 3.3E+20 | 0.12 | 8.5 |
| Result | 7.8E+20 | 1.3E+20 | 0.90 | 10.2 |

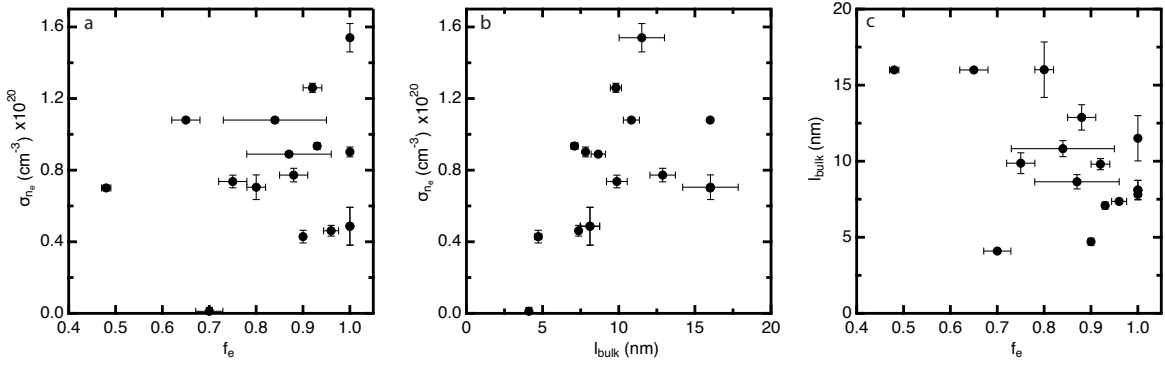

**Figure S4. No correlations in damping parameters.** Electron accessible volume fraction ($f_e$) and bulk mean free path ($l_{bulk}$) both contribute to intrinsic damping, while carrier concentration polydispersity ($\sigma_{n_e}$) contributes to heterogeneous broadening. These parameters show no correlation. Error bars are derived from fits to spectra of four dispersions of varying NC volume fraction ($f_v$).

**Table S2. Size series fit parameters and calculated properties.**

| | Nominal Diameter (nm) | 6 | 8 | 11 | 12 | 14 | 15 | 16 | 18 |
|---|---|---|---|---|---|---|---|---|---|
| Sample Details | $\mu_r$ (nm) | 3.31 | 3.94 | 5.83 | 5.96 | 7.01 | 7.71 | 7.99 | 8.83 |
| | $\sigma_r$ (nm) | 0.36 | 0.48 | 0.76 | 0.58 | 0.55 | 0.39 | 0.65 | 0.48 |
| | at% Sn | 5.42 | 5.54 | 5.03 | 5.36 | 5.03 | 4.79 | 4.74 | 4.61 |
| SDA w/ Floating fv | $w_p$ (cm$^{-1}$) | 12231 | 13875 | 13151 | 15240 | 14147 | 14256 | 14275 | 14812 |
| | $\Gamma$ (cm$^{-1}$) | 1561 | 1488 | 1360 | 1080 | 1027 | 1160 | 1169 | 1057 |
| | $n_e$ (cm$^{-3}$) | 6.7E20 | 8.6E20 | 7.7E20 | 1.0E21 | 8.9E20 | 9.1E20 | 9.1E20 | 9.8E20 |
| SDA w/ Fixed fv | $w_p$ (cm$^{-1}$) | 11487 | 13629 | 13136 | 15197 | 14126 | 14207 | 14241 | 14742 |
| | $\Gamma$ (cm$^{-1}$) | 2962 | 2071 | 1389 | 1254 | 1091 | 1275 | 1261 | 1274 |
| | $n_e$ (cm$^{-3}$) | 5.9E20 | 8.3E20 | 7.7E20 | 1.0E21 | 8.9E20 | 9.0E20 | 9.1E20 | 9.7E20 |
| HEDA | $\mu_{ne}$ (cm$^{-3}$) | 7.7E20 ±1.5E18 | 9.4E20 ±4.6E18 | 7.7E20 ±1.1E19 | 1.1E21 ±7.8E18 | 9.2E20 ±7.4E18 | 9.4E20 ±3.5E19 | 9.4E20 ±2.9E19 | 1.1E21 ±1.1E19 |
| | $\sigma_{ne}$ (cm$^{-3}$) | 7.0E19 ±1.6E18 | 1.1E20 ±1.0E18 | 1.3E20 ±2.5E18 | 7.0E19 ±6.9E18 | 7.7E19 ±3.8E18 | 1.1E20 ±1.4E18 | 8.9E19 ±1.2E18 | 7.4E19 ±3.5E18 |
| | $f_e$ | 0.49 ±0.00 | 0.65 ±0.03 | 0.92 ±0.02 | 0.80 ±0.02 | 0.88 ±0.03 | 0.84 ±0.12 | 0.87 ±0.1 | 0.75 ±0.03 |
| | $l_{bulk}$ (nm) | 17.0 | 17.0 | 9.8±0.4 | 16.0±1.8 | 12.9±0.8 | 10.8±0.5 | 8.7±1 | 9.9±0.7 |
| | $\Gamma$ (cm$^{-1}$) | 1530±6 | 1313±12 | 1021±10 | 973±28 | 877±23 | 914±16 | 997±6 | 944±23 |

**Table S3. Dopant concentration series fit parameters and calculated properties.**

| | at% Sn | 0 | 1.07 | 2.97 | 4.45 | 4.97 | 6.13 | 7.68 |
|---|---|---|---|---|---|---|---|---|
| Sample Details | $\mu_r$ (nm) | 8.91 | 9.56 | 9.31 | 9.23 | 9.46 | 8.80 | 8.72 |
| | $\sigma_r$ (nm) | 0.65 | 0.66 | 0.68 | 0.88 | 0.76 | 1.05 | 0.792 |
| SDA w/ Floating fv | $w_p$ (cm$^{-1}$) | 3805 | 8208 | 11750 | 14458 | 13959 | 15213 | 15928 |
| | $\Gamma$ (cm$^{-1}$) | 669 | 1081 | 1198 | 1076 | 1005 | 1361 | 1232 |
| | $n_e$ (cm$^{-3}$) | 6.5E19 | 3.0E20 | 6.2E20 | 9.3E20 | 8.7E20 | 1.0E21 | 1.1E21 |
| SDA w/ Fixed fv | $w_p$ (cm$^{-1}$) | 3640 | 8172 | 11750 | 14454 | 13965 | 15276 | 15942 |
| | $\Gamma$ (cm$^{-1}$) | 821 | 1139 | 1200 | 1090 | 979 | 1177 | 1176 |
| | $n_e$ (cm$^{-3}$) | 5.9E19 | 3.0E20 | 6.2E20 | 9.3E20 | 8.7E20 | 1.0E21 | 1.1E21 |
| HEDA | $\mu_{ne}$ (cm$^{-3}$) | 7.1E19 ±1.0E18 | 3.1E20 ±1.1E18 | 6.2E20 ±3.2E18 | 9.4E20 ±4.1E18 | 8.7E20 ±1.7E18 | 1.0E21 ±2.8E18 | 1.1E21 ±3.1E18 |
| | $\sigma_{ne}$ (cm$^{-3}$) | 1.2E18 ±2.2E18 | 4.3E19 ±3.5E18 | 9.4E19 ±1.8E18 | 4.6E19 ±3.0E18 | 4.9E19 ±1.1E19 | 1.5E20 ±7.9E18 | 9.0E19 ±2.7E18 |
| | $f_e$ | 0.70 ±0.03 | 0.90 ±0.00 | 0.93 ±0.00 | 0.96 ±0.02 | 1.00 ±0.00 | 1.00 ±0.00 | 1.00 ±0.00 |
| | $l_{bulk}$ | 4.1±0.0 | 4.7±0.2 | 7.1±0.2 | 7.4±0.1 | 8.1±0.6 | 11.5±1.5 | 7.8±0.1 |
| | $\Gamma$ (cm$^{-1}$) | 665±6 | 940±33 | 904±18 | 1016±13 | 921±44 | 828±54 | 1058±11 |

$\mu_r$ and $\sigma_r$ are the mean NC radius and its standard deviation, $w_p$ is the plasma frequency, $\Gamma$ is the damping frequency, $n_e$ is the free electron concentration, $\mu_{ne}$ and $\sigma_{ne}$ are the mean electron concentration and its standard deviation, $f_e$ is the fraction of electron accessible volume, and $l_{bulk}$ is the bulk mean free path.

**Table S4. Correlation parameters for doping series determined by significance value a = 0.05.**

|  | $r^2$ | y-int | SE y-int | slope | SE slope | t-score | t-value | Correlation |
|---|---|---|---|---|---|---|---|---|
| $D_{r_{NC}}$ | 0.50 | 6.81 | 0.94 | 0.45 | 0.20 | 2.22 | 0.034 | TRUE |
| $D_{n_e}$ | 0.02 | 8.15 | 4.16 | 0.27 | 0.90 | 0.31 | 0.385 | FALSE |
| $n_e$ | 0.95 | 1.6E20 | 6.7E19 | 1.4E20 | 1.4E19 | 9.77 | 0.000 | TRUE |
| $f_e$ | 0.72 | 0.80 | 0.04 | 0.03 | 0.01 | 3.57 | 0.006 | TRUE |
| $l_{bulk}$ | 0.65 | 4.44 | 1.09 | 0.72 | 0.24 | 3.06 | 0.011 | TRUE |
| $\Gamma$ | 0.38 | 789 | 78 | 29 | 17 | 1.74 | 0.066 | FALSE |
| $\Gamma_{bulk}$ | 0.00 | 570 | 70 | 0.89 | 15 | 0.06 | 0.477 | FALSE |
| $\Gamma_{surface}$ | 0.93 | 219 | 18 | 30 | 3.9 | 7.69 | 0.000 | TRUE |

**Table S5. Correlation parameters for size series determined by significance value a = 0.05.**

|  | $r^2$ | y-int | SE y-int | slope | SE slope | t-score | t-value | Correlation |
|---|---|---|---|---|---|---|---|---|
| $D_{r_{NC}}$ | 0.55 | 15.3 | 2.3 | 0.96 | 0.33 | 2.95 | 0.009 | TRUE |
| $D_{n_e}$ | 0.16 | 13.7 | 3.8 | -0.63 | 0.54 | 1.17 | 0.138 | FALSE |
| $n_e$ | 0.13 | 8.0E20 | 1.3E20 | 1.8E19 | 1.8E19 | 1.01 | 0.171 | FALSE |
| $f_e$ | 0.58 | 0.42 | 0.13 | 0.06 | 0.02 | 3.10 | 0.007 | TRUE |
| $l_{bulk}$ | 0.75 | 22.1 | 2.3 | 1.48 | 0.32 | 4.56 | 0.001 | TRUE |
| $\Gamma$ | 0.71 | 1642 | 148 | 88.3 | 21.3 | 4.15 | 0.002 | TRUE |
| $\Gamma_{bulk}$ | 0.77 | 109 | 64.4 | 45.1 | 9.3 | 4.88 | 0.001 | TRUE |
| $\Gamma_{surface}$ | 0.88 | 1533 | 132 | 133 | 19 | 7.02 | 0.000 | TRUE |

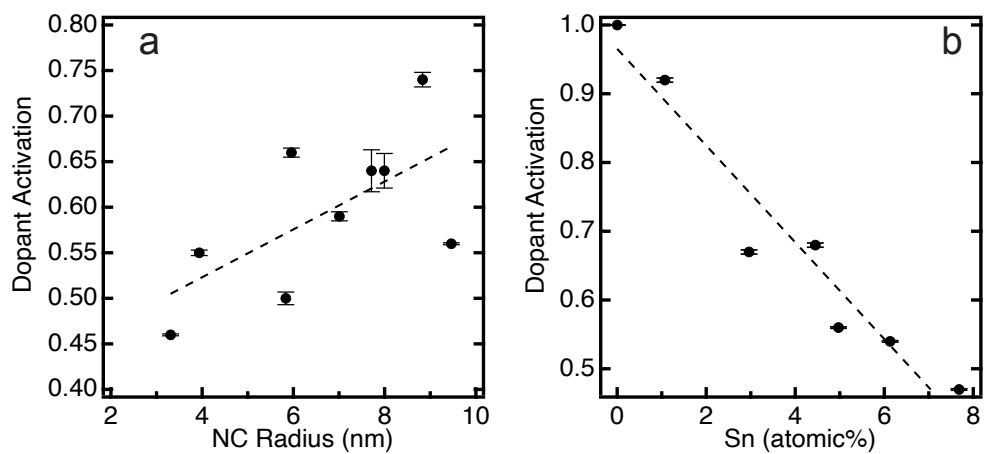

**Figure S5. Dopant activation.** After quantifying Sn atoms per NC from by inductively couple plasma-atomic emission spectroscopy (ICP-AES), see Reference 23 in the main text, the fraction of Sn dopants that contribute a free electron to the conduction band was plotted versus the NC radius (a) and Sn doping level (b).

**Table S6. Correlation parameters for dopant activation within the size and doping series determined by significance value a = 0.05.**

|  | $r^2$ | y-int | SE y-int | slope | SE slope | t-score | t-value | Correlation? |
|---|---|---|---|---|---|---|---|---|
| **Size Series** | 0.41 | 0.42 | 0.08 | 0.026 | 0.012 | 2.21 | 0.029 | TRUE |
| **Doping Series** | 0.93 | 0.97 | 0.04 | -0.07 | 0.01 | 8.42 | 0.000 | TRUE |

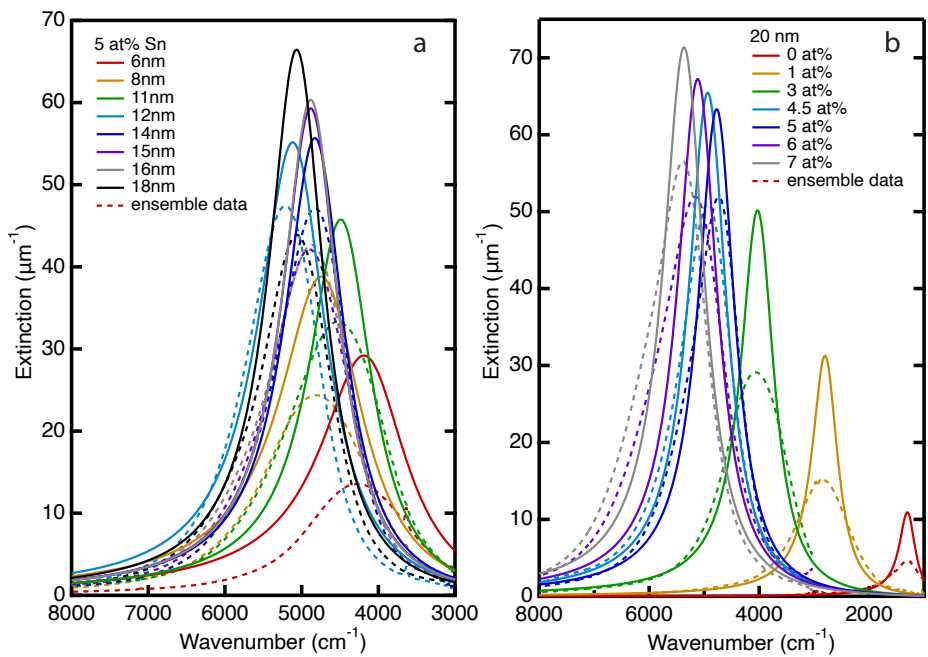

**Figure S6. SDA volume-normalized extinction compared to the ensemble spectrum.** Calculated extinction based of SDA fit results (solid) is significantly higher than that of the measured ensemble extinction (dashed) for samples in the size series (a) and the doping series (b).

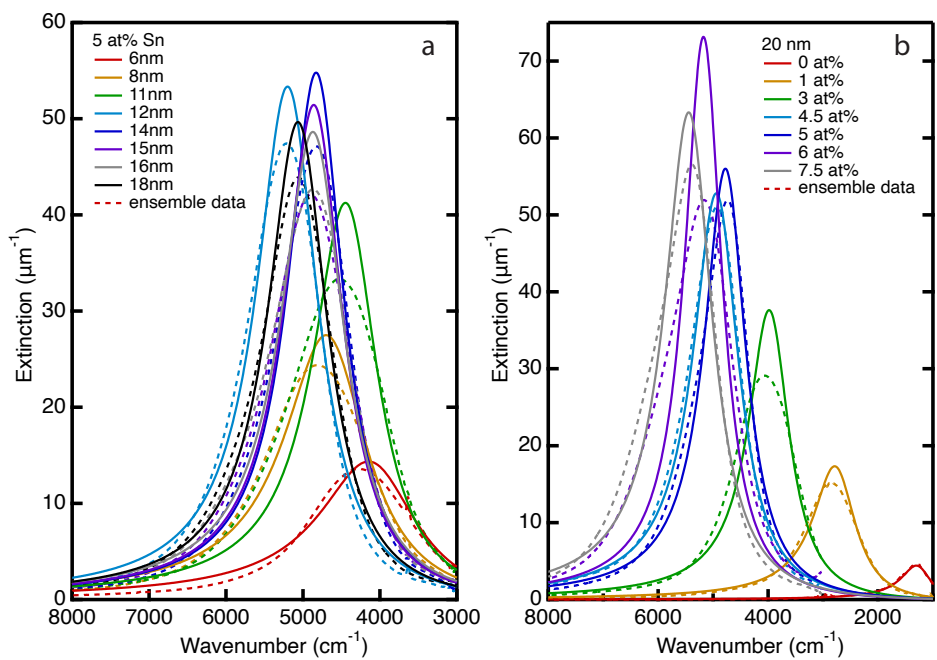

**Figure S7. HEDA-derived average NC extinction spectra compared to ensemble.** Average NC extinction coefficient (solid) is higher and shows a narrower lineshape than ensemble absorption coefficient (dashed) for samples in both the size series (a) and the doping series (b).

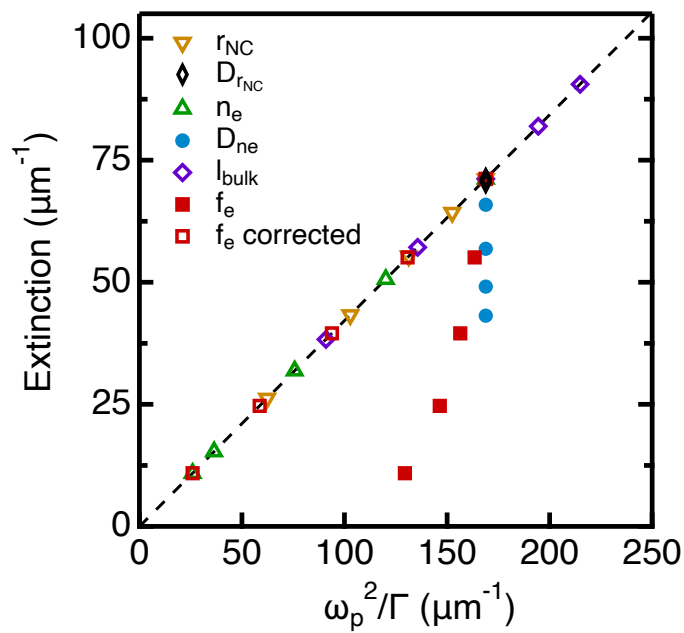

**Figure S8. Trends in optical extinction.** Peak volume-normalized extinction from all spectra plotted in Figure S2. The dashed line indicates the expected trend in extinction according to Mie scattering theory within the quasistatic limit for particles with $\omega_p \gg \Gamma$ (Equation 14). Ensemble heterogeneity and charge carrier surface depletion cause the peak extinction values to deviate. Using $f_e$ as a correction factor, the theory and calculated are in agreement.

**Table S7. Correlation parameters for size series determined by significance value a = 0.05.**

|  | $r^2$ | y-int | SE y-int | slope | SE slope | t-score | t-value | Correlation? |
|---|---|---|---|---|---|---|---|---|
| $Q_{enesmble}$ | 0.61 | 2.00 | 0.57 | 0.27 | 0.08 | 3.28 | 0.006 | TRUE |
| $Q_{NC}$ | 0.69 | 2.13 | 0.67 | 0.38 | 0.10 | 3.96 | 0.002 | TRUE |

**Table S8. Correlation parameters for doping series determined by significance value a = 0.05.**

|  | $r^2$ | y-int | SE y-int | slope | SE slope | t-score | t-value | Correlation? |
|---|---|---|---|---|---|---|---|---|
| $Q_{enesmble}$ | 0.67 | 2.30 | 0.42 | 0.28 | 0.09 | 3.18 | 0.010 | TRUE |
| $Q_{NC}$ | 0.82 | 2.52 | 0.47 | 0.48 | 0.10 | 4.80 | 0.002 | TRUE |

**Table S9. Correlation parameters determined by significance value a = 0.05.**

|  | $r^2$ | y-int | SE y-int | slope | SE slope | t-score | t-value | Correlation? |
|---|---|---|---|---|---|---|---|---|
| $\dfrac{Q_{NC}}{Q_{enesmble}}$ | 0.53 | 0.93 | 0.09 | 0.034 | 0.01 | 3.85 | 0.001 | TRUE |

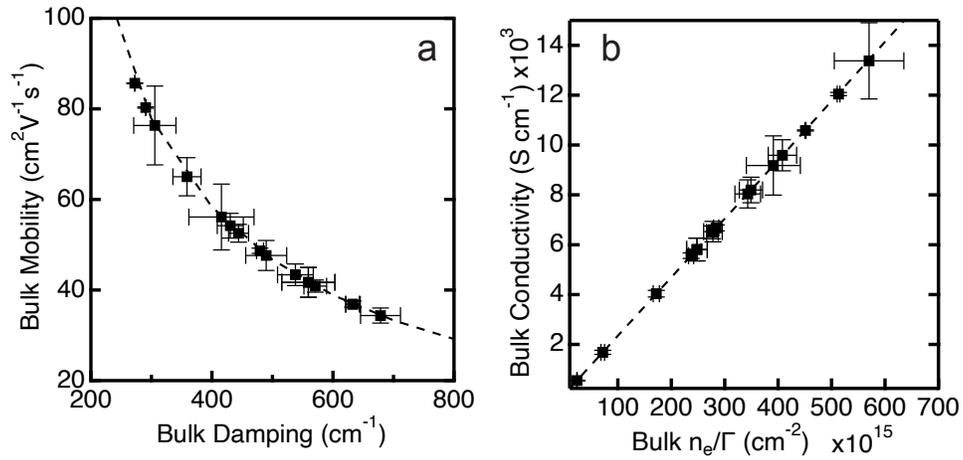

**Figure S9. Bulk ITO electronic properties.** After removing the effects of surface scattering, the bulk mobility (a) and bulk conductivity (b) are calculated from values extracted from the HEDA model for all ITO samples in this work. Dashed lines calculate the electronic parameters extrapolated beyond values extracted from our sample sets. Error bars indicate standard deviation in the results from HEDA fitting for four independently prepared dispersions of each sample.

# SI Text 1 MATLAB® code

Fitting File 1: "HEDA Fit"

```matlab
%% Initialize global variables for fitting

global epsilonNC epsilonSolvent pathLength r_range ne_range PD
epsilonNC=4.0;         % Dielectric background constant of nanocrystal [unitless] (ITO = 4)
epsilonSolvent=1.505^2; % Host/solvent Dielectric Constant [unitless]
lowFreqCutoff=1000;    % Low frequency cutoff x-axis [wavenumber 1/cm]
hiFreqCutoff=10000;    % High frequency cutoff x-axis [wavenumber 1/cm]
pathLength=0.05;       %Pathlength in cm

%% Load data
% The data should be in a text file and they should
% be formatted so that wavenumbers are in the first column and absorption
% values are in the second column
sample_name='name of text file here'
spectrum=dlmread('name of text file here.txt','\t',2,0);
wavenumbers=spectrum(:,1);  %load regular frequency values in cm-1
absorption=spectrum(:,2);   %load absorption values

% set fitting window
%set limits of fitting and grab indices
limits=find(wavenumbers>lowFreqCutoff&wavenumbers<hiFreqCutoff);
reducedFrequency=wavenumbers(limits);
reducedAbsorption=absorption(limits);%extract frequencies

global n_point p

n_point= 41
% p -- a vector of sample measurements:
%     p(1) -- radius stdev [nm]
%     p(2) -- radius mu_r [nm]
%     p(3) -- volume fraction [unitless]

p = [1.044  8.804  2.98E-05];

%% fitting
options=optimoptions('lsqcurvefit','Algorithm','trust-region-reflective','MaxFunEvals',1e20,'MaxIter',5e10,'TolFun',1e-14,'TolX',1e-15);
op.Display='on';
op.Plot=0;
op.ErrorsUnknown=1;  %set this to 1 if measurement uncertainties are unknown
op.MaxFunEvals=1e20;
```

```matlab
op.TolX=1e-20;              %Smallest step tolerance
op.TolFun=1e-20;
op.MaxIter=1e20;            %Maximum iterations possible
%op.FitUncertainty = [0 0 0 0 0 0];  %set to 1 for each parameter if uncertainty in fitted parameters is desired
LowerBound =  [1*(10^-3) 1*(10^-3) 0     0];
initialGuess= [6        1.3      0.8   1];
UpperBound =  [1*(10^3)  1*(10^3)  1     1.7];

paramsITO_ed=lsqcurvefit(@drude_sol,initialGuess,reducedFrequency,reducedAbsorption,LowerBound,UpperBound,options);
ne_mu=paramsITO_ed(1)*10^26;
ne_sigma=paramsITO_ed(2)*10^26;
dep=paramsITO_ed(3);
mfp=paramsITO_ed(4)*10;

Predicted=drude_sol(paramsITO_ed,reducedFrequency);
sample_name_fit=strcat(sample_name,'_fit');
results = paramsITO_ed
plot(reducedFrequency,reducedAbsorption,'b',reducedFrequency,Predicted,'r--')
FX=[reducedFrequency,reducedAbsorption,Predicted];

export_abs = reducedAbsorption;
export_fit = Predicted;
export_nu = reducedFrequency;
```

Fitting File 2: "drude_sol"

```matlab
function A=drude_sol(a,omega)
%   Input variables
%     omega -- frequency variable in cm^-1
%       a -- a vector of fit parameters:
%         a(1) -- ne average [m^-3]
%       a(2) -- ne st dev [m^-3]
%       a(3) -- electron accessible volume fraction [unitless]
%       a(4) -- bulk mean free path [nm]
% Output variable
% A -- absorbance of the layer

global epsilonNC epsilonSolvent pathLength n_point lower_limit upper_limit p r_range ne_range PD
vol_frac=p(3);
ravg=p(2);
rstdev=p(1);
```

```matlab
ne_mu = a(1)*10^26;
ne_sigma = a(2)*10^26;
dep = a(3);
mfp = a(4)*10;

r_range=(linspace(ravg-3*rstdev,ravg+3*rstdev,n_point))';
ne_range=linspace(ne_mu-3*ne_sigma,ne_mu+3*ne_sigma,n_point); %a(1)-3*a(2)
r_pdf=normpdf(r_range,ravg,rstdev)';
ne_pdf=normpdf(ne_range,ne_mu,ne_sigma);

abs_ensemble=zeros(length(omega),1);
PD=zeros(n_point,n_point);
T_PD=0;
V=0;
delr=(r_range(2)-r_range(1));
delne=(ne_range(2)-ne_range(1));

gamma=(((1.055*10^-34)*(3*pi^2)^(1/3)/(0.4*9.11*10^-31*3*10^10*2*pi)).*ne_range.^(1/3).*(1./(4/3*r_range*dep^(1/3)*10^-9)+1/(mfp*10^-9)));
omega_P=((ne_range)*(1.6*10^-19)^2/((8.85*10^-12)*(0.4*9.11*10^-31))).^(1/2)/(3*10^10)/2/pi;
omega_s=((10^23)*(1.6*10^-19)^2/((8.85*10^-12)*(0.4*9.11*10^-31))).^(1/2)/(3*10^10)/2/pi;

    for i = 1:n_point
    for j =1:n_point
            eshell=epsilonNC-omega_s^2./(omega.^2+1i*omega.*(gamma(i,j)));
            epsilonParticle=epsilonNC-omega_P(i)^2./(omega.^2+1i*omega.*(gamma(i,j)));
            e_eff_particle=eshell.*((epsilonParticle+2*eshell)+2*dep*(epsilonParticle-eshell))./((epsilonParticle+2*eshell)-dep*(epsilonParticle-eshell));
            sigA=4*pi*r_range(j)^3*2*pi*omega*sqrt(epsilonSolvent).*imag(( e_eff_particle-epsilonSolvent)./( e_eff_particle+2*epsilonSolvent));
            abs1=sigA;
            absc(i,j,:)=abs1;
            PD(i,j)=delr*delne*r_pdf(j)*ne_pdf(i);
            abs_ensemble=abs_ensemble+PD(i,j)*abs1; % adding onto the total abs
            T_PD=T_PD+PD(i,j);        %finding the total area of PDF for normalizing
            V=V+4/3*pi()*r_range(j)^3*PD(i,j);
    end
    end
T_PD;
A=abs_ensemble*vol_frac*pathLength/(V*log(10)*T_PD);
end
```